\documentclass[12pt]{spieman}  
\usepackage{amsmath,amsfonts,amssymb}
\usepackage{graphicx}
\usepackage{setspace}
\usepackage{tocloft}
\usepackage{color}

\title{Feasibility of non-invasive optical blood-glucose detection using overtone circular dichroism}

\author[a,b]{Brett H. Hokr}
\author[c]{Carlos E. Tovar}
\author[c]{Zhaokai Meng}
\author[c]{Georgi I. Petrov}
\author[a,c*]{Vladislav V. Yakovlev}
\affil[a]{Texas A\&M University, Department of Physics \& Astronomy, College Station, TX 77843}
\affil[b]{Engility Corp., San Antonio, TX 78228}
\affil[c]{Texas A\&M University, Department of Biomedical Engineering, College Station, TX 77843}

\cftpagenumbersoff{figure}
\cftpagenumbersoff{table} 
\begin{document} 
\maketitle

\begin{abstract}
Diabetes is one of the most debilitating and costly diseases currently plaguing humanity. It is a leading cause of death and dismemberment in the world, and we know how to treat it. Accurate, continuous monitoring and control of blood glucose levels via insulin treatments are widely known to mitigate the majority of detrimental effects caused by the disease. The primary limitation of continuous glucose monitoring is patient non-compliance due to the unpleasant nature of ``finger-stick" testing methods. This limitation can be largely, or even completely, removed by non-invasive testing methods. In this report, we demonstrate the vibrational overtone circular dichroism properties of glucose and analyze its use as a method of non-invasive glucose monitoring, capable of assuaging this trillion dollar scourge.
\end{abstract}


{\noindent \footnotesize\textbf{*}Vladislav V. Yakovlev,  \linkable{yakovlev@bme.tamu.edu} }

\begin{spacing}{1}   

\section{Introduction}
Diabetes is unquestionably one of the most serious epidemics society currently faces. In the United States alone, 29.1 million people suffer from the disease and it is the seventh leading cause of death. It is suspected that diabetes related deaths are significantly under reported by as much as a factor of two~\cite{AmericanDiabetesAssociation2014}. Additionally, it is estimated that the total cost of diabetes is \$245 billion dollars in just the United States~\cite{AmericanDiabetesAssociation2014}, and this neglects the life time of pain and suffering that diabetes patients endure through constant blood glucose testing. It has been well documented that these harmful effects of diabetes can be largely mitigated by effective insulin treatments~\cite{DiabetesControlandComplicationsTrialDCCTResearchGroup1993}. However, in order to accurately and safely provide insulin treatments, the blood glucose level must be monitored continuously~\cite{Boland2001}. Current monitoring techniques relying on ``finger-stick" methods cannot achieve continuous monitoring and suffer from non-compliance of patients due to the invasive nature of these techniques. Thus, it has been a long standing problem of science to develop a non-invasive and pain-free technique to monitor blood glucose levels in diabetic patients~\cite{Malin1999,Arnold2005,Larin2003,Tamada1999,Burmeister1999,Khalil1999}.

There are many proposed techniques for non-invasive glucose monitoring, the vast majority of which are optical in nature. This is in large part due to the fact that they utilize nonionizing radiation, do not typically require consumable reagents, and are possible to do in real time. These techniques include near-infrared spectroscopy~\cite{Malin1999}, monitoring changes to the scattering properties of inter-cellular fluids~\cite{Maier1994}, Raman spectroscopy~\cite{Berger1997a}, polarimetry~\cite{Cameron1999}, photoacoustics~\cite{MacKenzie1999}, and many more~\cite{McNichols2000}. All of these methods suffer from the same core problem, many biological molecules have similar effects and are often in greater concentration than glucose. Discriminating the effect of glucose, from the effect of these other molecules has proved difficult and has consequently been termed the so called ``calibration problem"~\cite{McNichols2000}. Other methods rely on secondary indicators such as aqueous humor glucose or interstitial fluid glucose, however, because of the time lag in response time from these indicators, additional variables have to be factored into the "calibration problem" in order to obtain reliable data from the spectrum.

Glucose is a chiral molecule, with the overwhelming majority of glucose molecules occurring in nature being that of the D-isomer. As a result, the glucose molecule, as with any other chiral molecule, exhibits a difference in the absorption strength of right-hand circular and left-hand circular light, known as circular dichroism~\cite{Berova2000}. Circular dichroism exhibited by vibrational energy levels is known as vibrational circular dichroism to distinguish it from circular dichroism on electronic transitions. The vibrational circular dichroism of glucose has been studied on the fundamental vibrational transitions extensively. The spectral ranges 900 to 1200~cm$^{-1}$~\cite{Bose1999,Polavarapu2000}, 1175 to 1575 cm$^{-1}$~\cite{Tummalapalli1988}, and 2750 to 3050 cm$^{-1}$~\cite{Taniguchi2007} have been reported in the literature. These results are of little interest for non-invasive glucose detection due to the strong absorption of water in this spectral region severely limiting the penetration depth. To avoid this critical limitation, we must move towards overtone absorption transitions. Another significant advantages of working in this spectral region is a much larger selection of affordable and reliable sources and detectors, making the possibility of a commercially viable device more realistic.

In this report, we propose and demonstrate preliminary evidence for the non-invasive detection of blood glucose using overtone vibrational chiral dichroism~\cite{Abbate2000,Abbate2009}. This method has the potential to separate the presence of glucose from its surroundings, avoiding the ``calibration problem", while circumventing the huge absorption of water in the fundamental vibrational region of the spectrum, allowing for clinically relevant penetration depths in biological tissue. In addition to this, the polarization of light is preserved through tens of scattering events~\cite{Jarry1998}, giving this technique the ability to penetrate reasonably deep into tissue, even in the presence of the large amounts of scattering in biological tissues.

\section{Materials and Methods}
\begin{figure}[tb]
\centering
\includegraphics[width=0.9\textwidth]{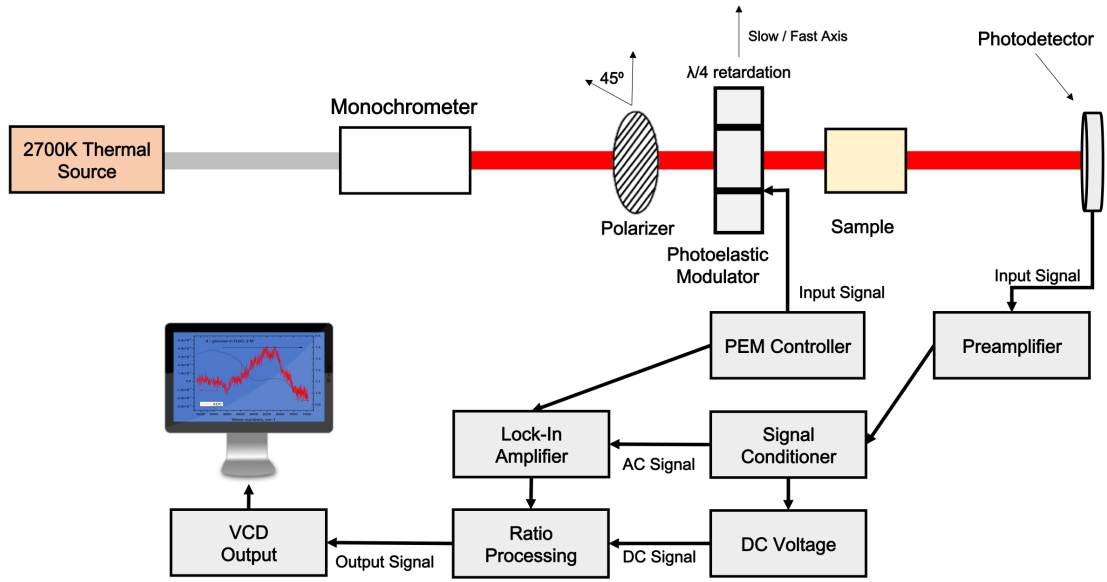}
\caption{Schematic diagram of the experimental setup.}
\label{fig:Schematic}
\end{figure}

The experimental setup is depicted in Fig.~\ref{fig:Schematic}. To measure the vibrational overtone circular dichroism spectrum, a narrow spectral beam was obtained from a stabelized tungsten lamp (Thorlabs; SLS201) by passing it through a monochrometer (Horiba; TRIAX 320) with a 600 ln/mm infrared grating blazed at 1000~nm. The slit on the monochrometer was set to allow a spectral resolution of 3~nm across the spectral range from 1.10~$\mu$m to 1.30~$\mu$m, corresponding to the second CH stretch overtone region of the spectrum. The power output of the source at this resolution was between 5 and 8 $\mu$W across this spectral range. The light was then passed through a 45$^\circ$ Glan-Taylor polarizer (Lambda Research Optics; CGTP-15) followed by a photoelastic modulator (PEM) (Hinds Instruments; I/FS50 head with PEM-90 controller) that modulates the beam between right-hand circular and left-hand circular polarization at a frequency of 100.22~kHz. The beam is then passed through the sample and detected by a switchable gain amplified InGaAs photodiode (Thorlabs; PDA20CS).

The signal from the photodiode is split into an AC component at the frequency of the PEM which is due to the change in absorption from left-hand circular to right-hand circular and a DC component that is a measure of the total absorption. The ratio of these two gives the difference between the absorbance for left-hand circular and right-hand circular. In this case, the AC signal is proportional to the difference in the transmissions as given by the Beer-Lambert law of each polarization, $e^{-A_R} - e^{-A_L}$, and the DC signal is proportional to the average of the two transmissions, $(e^{-A_R} + e^{-A_L})/2 \approx e^{-A_R}$. Here we have made use of the fact that the amount of circular dichroism is small, typically on the order of $10^{-6}$. The ratio of these signals becomes
\begin{equation}
\mathrm{\frac{AC}{DC}} = \frac{e^{-A_R} - e^{-A_L}}{e^{-A_R}} = 1-e^{-(A_L-A_R)} \approx A_L - A_R = \Delta A.
\end{equation}
To obtain the molar circular dichroism we have
\begin{equation} \label{eq:delta_epsilon}
\Delta \epsilon = \frac{\Delta A}{C L}
\end{equation}
where $C$ is the concentration, and $L$ is the path length of the sample. To obtain the absorption of the sample, a reference measurement is required to determine the transmission of the system in the absence of the absorber. We can write the absorbance of the target in terms of the DC signal from the sample, and the DC signal from the reference sample, $\mathrm{DC_{ref}}$,
\begin{equation}
A = \log \left (\mathrm{\frac{DC}{DC_{ref}}} \right ).
\end{equation}
Similar to \eqref{eq:delta_epsilon}, the molar absorption coefficient is obtained by dividing by the molar concentration and the sample length. 

\begin{figure}[tb]
\centering
\includegraphics[width=0.6\textwidth]{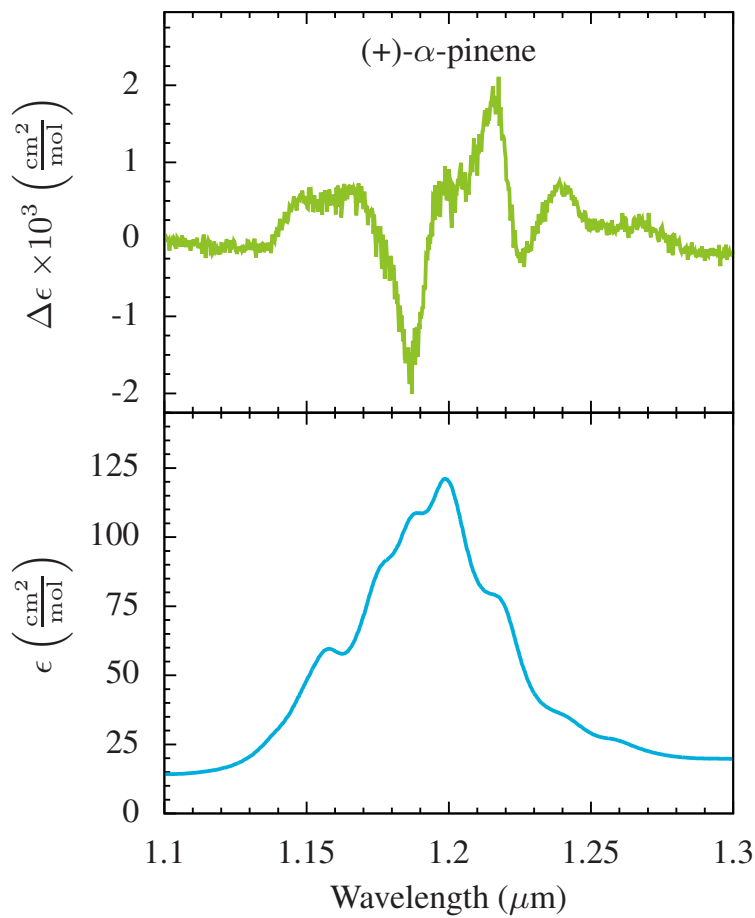}
\caption{Overtone vibrational circular dichroism (top) and absorption (bottom) spectrum for $(+)$-$\alpha$-pinene.}
\label{fig:Pinene}
\end{figure}

To validate the existing experimental setup, a sample of undiluted $(+)$-$\alpha$-pinene (Sigma Aldrich; CAS 7785-70-8) was measured in a 1 cm Infrasil low-birefringence 10 mm cuvette (Starna Inc.), and compared to previously reported results~\cite{Guo2006}. Thus, an empty cuvette was used as the reference sample for the absorption calculations. This potentially introduces error into the absorption numbers by neglecting Fresnel reflections from the cuvette walls; however, for the purpose of verifying the validity of the setup, this is acceptable and does not, in principle, effect the circular dichroism measurement. To obtain the molar absorption and circular dichroism values, $6.3$~M was used as the molar concentration for the pure $(+)$-$\alpha$-pinene. Both the absorption and circular dichroism spectra for $(+)$-$\alpha$-pinene measured by this setup are shown in Fig.~\ref{fig:Pinene}. Our results are in good agreement with previous results reported in the literature~\cite{Guo2006}.

\section{Results and Discussion}
\begin{figure}[tb]
\centering
\includegraphics[width=0.6\textwidth]{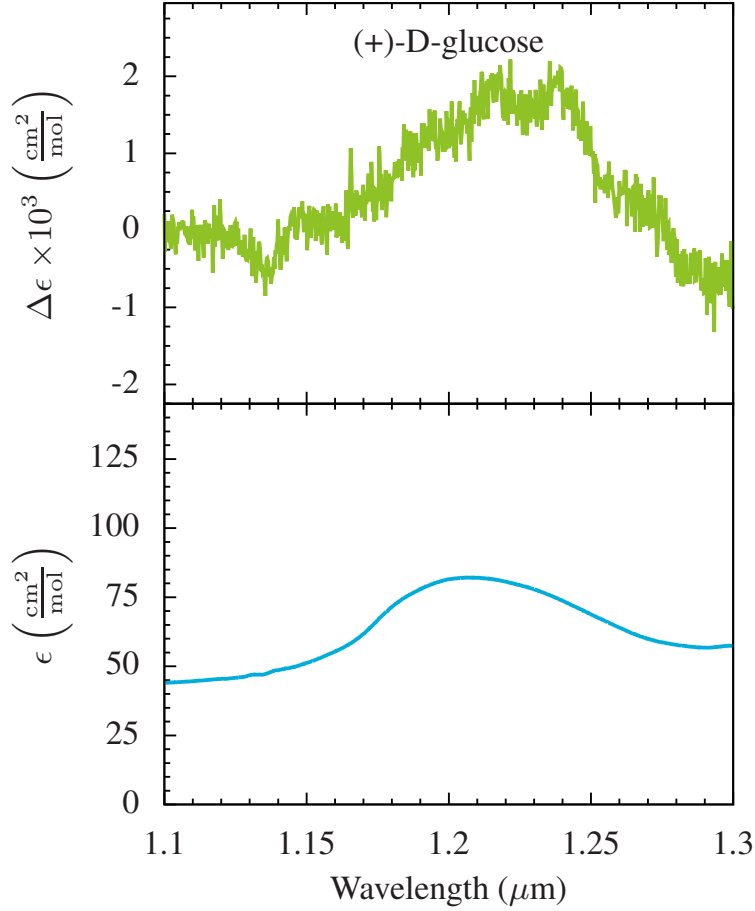}
\caption{Overtone vibrational circular dichroism (top) and absorption (bottom) spectrum for $(+)$-D-glucose.}
\label{fig:Glucose}
\end{figure}

To measure the overtone circular dichroism of glucose, a solution of 2~M (+)-D-glucose (Sigma Aldrich; CAS 50-99-7) was dissolved in heavy water (D$_2$O) and left overnight to ensure homogenation. A cuvette filled with pure D$_2$O was used as the absorption reference. The results are shown in Fig.~\ref{fig:Glucose}, and show a measurable amount of circular dichroism present in a region of the spectrum that has substantial penetrating ability in tissue; thus, making non-invasive measurements of blood glucose by near-infrared circular dichroism possible.

This setup can be further improved by using a tunable laser source, such as an external cavity tunable diode laser, in place of the tungsten light source and monochrometer, which can only deliver micro Watts of power. In contrast, a laser source would be able to provide powers limited only by the damage threshold of the tissue, which in this wavelength range can be on the level of Watts, depending on the spot size used~\cite{Bixler2014a,Bixler2015a}. This five orders of magnitude increase in available power will dramatically increase the signal to noise of the measurements while simultaneously decreasing the time needed for the measurement. In addition to the increase in power, a tunable laser source would offer much higher spectral resolution than the thermal light source, allowing for improved specificity, because small changes in the spectrum would be more easily observed.

In order for a non-invasive vibrational circular dichroism measurement to be possible, two major issues must be addressed: first, the light on the target must maintain a decent degree of polarization through tissue, and second, an efficient detection method must be found capable of detecting glucose levels with sensitivity of at least 5.6~mM (100~$ \mathrm{mg/dL} $) and an accuracy exceeding about 1~mM (18~$ \mathrm{mg/dL} $) to make the results clinically relevant~\cite{WHO2006}

To investigate the issue of non-perfect initial polarization, we must look at the error such a situation introduces into the measurements. To start, lets assume that our initial right-hand circular beam is comprised of a fraction of right-hand circular light, $r_R$, and some left-hand circular light, $l_R$, such that
\begin{equation}
I_{0,R} = I_0 (r_R + l_R).
\end{equation}
Upon passing through a sample that exhibits circular dichroism, the right-hand circular light is attenuated by $e^{-A_R}$ and the left by $e^{-A_L}$, thus the transmitted light is
\begin{equation}
I_R = I_0 \left ( r_R e^{-A_R} + l_R e^{-A_L} \right ).
\end{equation}
Using a similar definition for mostly left-hand circular light, making the assumption that both beams have the same degree of polarization $\delta = r_R-l_R = l_L-r_L$, and following the same analysis used above for purely polarized beams, we find
\begin{equation}
\mathrm{\frac{AC}{DC}} = \delta \Delta A.
\end{equation}
Thus, any kind of depolarization of the incident beam, prior to interaction with the molecules of interest constitutes a linear decrease in the measured circular dichroism, and perhaps most importantly, does not intrinsically have a wavelength dependence which would alter the shape of the spectrum. Strictly speaking, the amount of depolarization due to scattering in tissue will be wavelength dependent, but in the near infrared wavelength range, the anisotropy, $g = \langle \cos(\theta)\rangle$ of scattering from tissue is in the range 0.6 to 0.9~\cite{Troy2001,Ahmad2011,Jacques2013}. This indicates that scattering is well into the Mie regime and is due to larger scattering centers. In this regime, incident polarization is largely maintained from one scattering event to the next and reasonable degrees of polarization are possible millimeters deep in tissue~\cite{Jarry1998,Doronin2014}. This means that significant penetration depths, millimeters deep, are accessible in biological tissue with near infrared circular dichroism. This could be a potential source of error for \emph{in vivo} measurements and further studies in this direction are certainly needed to determine if this error from depolarization is tolerable or will require calibration to remove. There are analytical models which have been proposed that could better model the effect of scattering and reduce, or remove this error entirely~\cite{Pham2012,Liao2013}.

Another major challenge of making measurements \emph{in vivo} is that scattering from biological tissue complicates detection as it is no longer as simple as a transmission measurement. There are several techniques that are commonly used to circumvent this problem, such as diffuse reflectance spectroscopy~\cite{Malin1999}. However, these methods suffer from a few problems that make them unattractive for applications to vibrational circular dichroism measurements. First, these are typically more of a bulk measurement where you are sampling a large volume of tissue. This may or may not be a bad thing, but it will certainly introduce a lot of additional artifacts into the measurements that would not be present if you could restrict your measurement to a blood vessel. Another problem with these bulk techniques is that they more strongly rely on the assumption that the scattering properties of tissue are the same for right-hand circular as left-hand circular down to $10^{-6}$ level effects. Given the inhomogeneous and complicated nature of scattering in biological tissue this seems likely to not be the case. 

A better option for accurately measuring the vibrational circular dichroism spectrum of glucose \emph{in vivo} is photoacoustic detection~\cite{Zhang2006}. Due to the much lower scattering of acoustic waves in tissue, you can determine where in the tissue your signal is coming from, making it possible to isolate the signal from blood~\cite{Petrova2005}. In principle, differences in scattering can still effect these measurements, but their effect should be greatly decreased because only the photon's trip to the absorbing molecule plays a role, instead of the full path in the bulk measurements listed previously. Additionally, spurious noise from environmental effects can be greatly reduced through high-speed lock-in detection~\cite{Zhao2014,Wilson2015}.

While an accurate, non-invasive glucose sensor remains illusive, the increased chemical specificity and potential penetration depth of vibrational circular dichroism in the near-infrared regime makes it a viable candidate that warrants significant further study. In this report, the vibrational circular dichroism spectrum for the second C-H stretch overtone has been presented. Furthermore, we have explored the feasibility of using this methodology as a non-invasive \emph{in vivo} technique for the continuous monitoring of blood glucose levels.

\acknowledgments 
We acknowledge the support of the National Science Foundation (CBET award 1250363, DBI awards 1455671 and 1532188, and ECCS award 1509268). BHH would like to acknowledge a graduate fellowship from the Department of Defense Science, Mathematics and Research for Transformation (SMART) fellowship program.



\end{spacing}
\end{document}